\documentclass[%
reprint,
twocolumn, 
showpacs,
 amsmath,amssymb,
aip,
jcp,
]{revtex4-1}
\usepackage{graphicx}% Include figure files
\usepackage{dcolumn}% Align table columns on decimal point
\usepackage{bm}% bold math
\usepackage[utf8]{inputenc}

\usepackage{etex}
\usepackage{default}

\usepackage{graphicx}
\usepackage{color}
\usepackage{colortbl}
\usepackage{xcolor}
\usepackage{calc}

\usepackage{amsfonts}
\usepackage{amssymb}
\usepackage{amsmath}
\usepackage{wasysym}

\usepackage{float}
\usepackage{epstopdf}
\usepackage{nicefrac}

\usepackage{tcolorbox}% http://ctan.org/pkg/tcolorbox
\usepackage{marvosym}

\usepackage[makeroom]{cancel}

\usepackage[vcentermath]{youngtab} % for Young tableaux
\usepackage{tikz}

\usetikzlibrary{trees}
\usetikzlibrary{calc}
\usetikzlibrary{positioning}
\usetikzlibrary{shapes.misc}
\usetikzlibrary{decorations.pathreplacing}

\usetikzlibrary{shapes.arrows} 

\tikzset{
  solid node/.style = {circle, draw, inner sep = 3, fill = black},
  right angle/.style = {grow = 300},
  left angle/.style = {grow = 240},
  short/.style = {level distance = 1cm, },
  long/.style = {level distance = 2cm},
  left up/.style = {grow = 120},
  right up/.style = {grow = 60}
}

\usepackage{rotating}
\usepackage{mathtools}
\usepackage{hyperref}

\usepackage{units}
\usepackage{tabularx}
\usepackage{pgf}
\usepackage{pgfplots}
\usepackage{geometry}
\usetikzlibrary{calc}
\usepackage{relsize}
\usepackage{longtable}
\usepackage{dcolumn}
\usepackage{colortbl}

{%
   \end{list}%
}%

\newlength{\marginwidth}
\newcommand*{\dashfill}{\leavevmode\cleaders\hbox{-}\hfill\kern0pt}

\newcommand*{\midhrulefill}{
\leavevmode
\cleaders\hbox to 1ex{\raisebox{.5ex}{\rule{1ex}{.4pt}}}\hfill\kern0pt
}

\newcommand{\ket}[1]{{\left| #1 \right>}}

\newcolumntype{L}{>{$}l<{$}} % math-mode version of "l" column type
\newcolumntype{C}{>{$}c<{$}} % math-mode version of "l" column type

\tikzset{site/.style={circle,very thick,draw,opacity=0.5,inner sep=2pt}}

\newlength{\latticesep}
\definecolor{doublecolor}{RGB}{200 0 0}
\definecolor{singlecolor}{RGB}{0 200 0}
\tikzset{lattline/.style={-,thick}}
\tikzset{fermi/.style={rounded corners}}
\tikzset{line/.style={very thick}}
\tikzset{arrow/.style={->,very thick}}
\tikzset{box/.style={fill=white,rounded corners=1pt, opacity = 0.95, text opacity = 1.0}}

\def \beq {\begin{eqnarray}}
\def \eeq {\end{eqnarray}}

\def \Schrodinger {{Schr\"{o}dinger }}

\begin{document}

%\preprint{APS/123-QED}

\title{Unbiasing the initiator approximation in Full Configuration Interaction Quantum Monte Carlo} 
% Force line breaks with \\

\author{Khaldoon Ghanem}
\affiliation{%
Max Planck Institute for Solid State Research, Heisenbergstr. 1, 70569 Stuttgart, Germany
}
\author{Alexander Y. Lozovoi}
\affiliation{%
Max Planck Institute for Solid State Research, Heisenbergstr. 1, 70569 Stuttgart, Germany
}
\affiliation{
Dept of Physics, King's College London, Strand, London WC2R 2LS, United Kingdom
}%
\author{Ali Alavi \thanks{\mailto:alavi@fkf.mpg.de}}
\email{a.alavi@fkf.mpg.de}
\affiliation{%
Max Planck Institute for Solid State Research, Heisenbergstr. 1, 70569 Stuttgart, Germany
}
\homepage{https://www.fkf.mpg.de/alavi}

\affiliation{%
Dept of Chemistry, University of Cambridge, Lensfield Road, Cambridge CB2 1EW, United Kingdom
}%

\date{\today}% It is always \today, today,

\begin{abstract}
We identify and rectify a crucial source of bias in the initiator FCIQMC algorithm. 
Non-initiator determinants (i.e. determinants whose population is below 
the initiator threshold) are subject to a systematic {\em undersampling} bias, which in large systems leads to a 
bias in the energy when an insufficient number of walkers is used.  
We show that the acceptance probability ($p_{acc}$), that a non-initiator determinant has its spawns accepted, 
can be used to unbias the initiator bias, in a simple and accurate manner, by reducing the applied shift to the non-initiator proportionately to $p_{acc}$. This modification preserves the property that in the large walker limit, when $p_{acc}\rightarrow1$, the unbiasing procedure disappears, and the initiator approximation becomes exact. We demonstrate that this algorithm shows rapid convergence to the FCI limit 
with respect to walker number, and furthermore largely removes the dependence of the algorithm on the initiator threshold, enabling highly accurate results to be obtained even with large values of the threshold.  
This is exemplified in the case of butadiene/ANO-L-pVDZ and benzene/cc-pVDZ, correlating 22 and 30 electrons in 82 and 108 orbitals respectively. In butadiene $5\times 10^7$ and in benzene $10^8$ walkers suffice to obtain an energy to within a milli-Hartree of the CCSDT(Q) result, in Hilbert spaces of $10^{26}$ and $10^{35}$ respectively. Essentially converged results require $\sim 10^8$ walkers for butadiene and $\sim 10^9$ walkers for benzene, and lie slightly lower than CCSDT(Q). Owing to large-scale parallelisability, these calculations  
can be executed in a matter of hours on a few hundred processors. 
The present method largely solves the initiator-bias problems that the initiator method suffered from 
when applied to medium-sized molecules.  
\end{abstract}

%\begin{description}
%Secondary publications and information retrieval purposes. TODO
%\end{description}

%\keywords{Similarity Transformation}%Use showkeys class option if keyword
\maketitle

%\tableofcontents

\section{\label{sec:intro}Introduction}

The FCIQMC algorithm \cite{original-fciqmc} is a projective QMC method designed to solve the electronic Schrödinger eigenvalue problem in a given basis set at the full-configuration interaction level.
It is based on a population dynamics of a set of positive and negative walkers,  the master 
equation of which is derived by interpreting the imaginary-time Schrödinger equation as a first-order kinetic equation. 
In the limit of a large number of walkers under steady-state conditions, 
the population dynamics samples the exact  fermionic ground-state wavefunction. The algorithm is highly flexible, being generalisable to a number of different types of problems, including sampling excited states of the same symmetry as the ground state \cite{excited-fciqmc}, to complex wavefunctions appropriate for solids \cite{Nature2013}, to the real-time domain for spectroscopic applications \cite{Guther2018}, to Jastrow-factorised explicitly-correlated wavefunctions \cite{Luo2018,Dobrautz2019,Cohen2019}, and to spin-adaptation via the graphical Unitary Group approach \cite{Dobrautz2019_2}. There are two forms of the algorithm, a ``full''  formulation in which the Hamiltonian is applied in unconstrained form, and an ``initiator'' approximation (i-FCIQMC) \cite{initiator-fciqmc} in which a constraint is applied to the Hamiltonian, to be discussed in detail later.
%dynamically setting to zero those Hamiltonian matrix elements which connect a non-initiator determinant with an empty determinant. 

In its full form, FCIQMC converges without bias onto the ground-state eigenvector of a Hamiltonian, assuming it to be non-degenerate (degenerate or near-degenerate cases are treatable via the excited-state approach). However, 
the full version of FCIQMC requires a minimum number of walkers to do so - simulations with insufficient numbers of walkers are unable to stably converge onto the exact solution. This 
number is both system  and basis dependent, and is usually found to be smaller than the size of the Hilbert space, implying
a lower memory requirement compared to iterative exact diagonalisation. 
However it is also found to scale with the size of the Hilbert space (for example as the number of electrons is increased), 
making it impractical for many systems of interest.  In other words, the FCIQMC algorithm has an exponential scaling memory requirement, similar to that of iterative methods such as the Lanczos or Davidson algorithms.  

The instability observed in the sub-minimum walker regime of the full FCIQMC algorithm is a manifestation of the sign-problem in this method, 
which has been discussed by Spencer et al \cite{Spencer2012} in terms of competition with the ground state of a different (sign-problem-free)  
Hamiltonian with off-diagonal elements given by $-|H|$, the latter dominating in the sub-critical walker regime. 
In essence, an insufficient number of walkers means that the walker annihilation events of the algorithm do not occur with sufficient
frequency, and the correct  permanently established  sign-structure of the CI coefficients cannot emerge from the random dynamics of the method. 
In fact determinants which are not permanently occupied, but are only visited occasionally,   
fluctuate in sign as they are visited by walkers of either sign. Such sign-fluctuating determinants are a source of sign-incoherent noise: 
their progeny also fluctuate in sign, thereby propagating this noise exponentially. In order to 
prevent this, in the ``initiator'' method a constraint is placed on the spawning step of the algorithm.
The instantaneous distribution of walkers is divided into two (dynamically evolving) sets: those walkers which reside on determinants populated by more than a certain number $n_a$ of walkers (typically set to 3)  
are deemed to be ``initiators''. Such determinants are deemed to have the correct sign, and they are allowed to freely spawn progeny on connected determinants, as dictated by the Hamiltonian. 
Those walkers which reside on determinants occupied by less than or equal to $n_a$ walkers are designated as `non-initiators'. They are allowed to spawn progeny only on already occupied determinants (initiators or non-initiators). In other words, in 
the initiator approximation certain off-diagonal Hamiltonian matrix elements of low-amplitude determinants are dynamically discarded. (The word dynamical is used to emphasize that, as the distribution of walkers changes from iteration to iteration, the discarded part of the Hamiltonian also changes. It is not a fixed set, determined a priori by a selection criterion). It is found that with this modification, {\em stable simulations can be performed at any walker number (however small)}, thus obviating the memory bottleneck of the full algorithm. However, this comes
at the cost of a systematic bias in the computed energy. This ``initiator'' bias can be made arbitrarily small by increasing the walker number, and indeed the algorithm is designed
to revert to the `full' (i.e. exact) algorithm in the limit of large number of walkers.  In practice, for systems up to about ~20 electrons, convergence can be achieved with respect to walker number, well before memory requirements have become impractical. However, as the system size grows, it has been found that the convergence with respect to walker number slows down, such that it becomes practically impossible to converge to the exact FCI limit. 

In this paper we show that the initiator bias can be easily rectified as the simulation proceeds, enabling convergence to the FCI limit with relatively small number of  walkers, several orders of magnitude fewer than that required by the initiator method or the full FCIQMC method.  The methodology yields not only near-exact FCI-level energies, but also the reduced density matrices can be obtained via replica method \cite{Overy2014}, enabling property calculation. 
The latter will be the subject of a forthcoming publication.

Recently Blunt \cite{Blunt2018} has proposed a perturbative correction to estimate the initiator error with respect to a variational
estimate of the i-FCIQMC energy obtained from the reduced density matrices. This method is in the spirit of the Epstein-Nesbet PT2 correction \cite{Garniron2017,Sharma2017} of the 
selected CI methods such as CIPSI \cite{Huron1973}, Heat-bath CI \cite{Holmes2016}, and other adaptive methods \cite{Evangelista2014,Tubman2016,Schriber2016,Liu2016}. These methods can be used to extrapolate to the 
 $E_{PT2}\rightarrow 0$ limit, thereby producing estimates of the FCI energy. However they 
crucially rely on efficient hybrid stochastic means to obtain the perturbative energy corrections, and do not easily yield 
corrections to the {\em wavefunctions} without substantial computational overhead. This makes the calculation of properties at the corresponding level of accuracy (i.e. approaching FCI) very difficult.

Ten-no \cite{Tenno2017}  has discussed the initiator approximation in terms of size inconsistency error, and has proposed several ways to mitigate this via  coupled electron pair type approximations.  The method proposed here has a resemblance to these concepts, but the form of the correction is different, being adapted to each non-initiator determinant rather than prescribed, and vanishes in the larger walker limit and thereby ensuring exactness in that limit.  

Incremental many-body expansions (MBE) of FCI \cite{Eriksen2017, Eriksen2019, Zimmerman2017} offer an alternative approach to the FCI problem, but do involve a large number of sub-space CASCI diagonalisations, which for large systems may become too large for deterministic diagonalisation. In such problems the method to be discussed below could be used in conjunction with MBE-type methods to alleviate those bottlenecks. 

Another highly promising approach that could benefit from the present methodology is the cluster-analysis-driven (CAD) FCIQMC methodology of Piecuch and coworkers \cite{Deustua2018,Deustua2019}, who solve the CCSD amplitude equations in the presence of the $T_3$ and $T_4$ amplitudes extracted from FCIQMC propagations. If the $T_3$ and $T_4$ amplitudes are exact, the resulting energies from these equations are also exact (i.e. equivalent to FCI). Piecuch et al have demonstrated this for small systems such as the water molecule, yielding exact energies from information derived from relatively short FCIQMC propagations. The present methodology may be a route to accurate $T_3$ and $T_4$ amplitudes at an affordable cost for larger systems, and would result in a very powerful combination of Coupled Cluster theory and FCIQMC. 
   
The structure of this paper is as follows. We first review the initiator FCIQMC method and identify a source of bias which results from the initiator approximation. We then discuss a method which we call the {\em adaptive-shift method} to unbias for this error on the fly, and discuss its implementation. Next we show how this methods works in the case of butadiene and benzene in double-zeta basis sets. We end with some concluding remarks on future perspectives.  
   
\section{The Initiator and Adaptive shift methods}   
 We begin by reviewing the main concepts behind FCIQMC and i-FCIQMC algorithms. 
 The imaginary-time \Schrodinger equation for the wavefunction $\Psi$ is:
 \beq
   -\partial_t \Psi=(\hat{H}-E)\Psi=0   \label{itse}
 \eeq
 $\Psi$ is expanded in an FCI basis:
 \beq
 \Psi=\sum_i c_i\ket{D_i}
 \eeq    
 where the coefficients $c_i$ are to be determined to achieve the stationarity condition implied by Eq.~\ref{itse}. In FCIQMC in its original formulation, a distribution of $N_w$ signed walkers 
 $\{i_1, i_2,...,i_\gamma,.., i_{N_w}\}$) of unit amplitude ($\{s_\gamma=\pm 1\},  N_w=\sum_\gamma |s_\gamma|)$ is invoked, such that $c_i$ coefficients are given via the relation
 \beq
 c_i\propto \sum_\gamma s_\gamma \delta(i_\gamma-i)\equiv N_i   \label{deltafunc}
 \eeq
In a subsequent development of FCIQMC \cite{Umrigar2012,Blunt2015}, the weights $s_\gamma$ were generalised to floating point 
numbers with the condition $|s_\gamma|\ge s_{cut}$, where $s_{cut}$ denotes the minimum amplitude of a walker, here taken to be 1. This modification allows for a much finer resolution of the instantaneous wavefunction to be achieved without permanent storage of 
excessively small determinant weights, and leads to faster convergence with population, and smoother convergence with imaginary time. This version of the algorithm, as implemented in the NECI code \cite{NECImolphys2014} with floating point walker weights, is the one we will employ in this study. 
 
 According to Eq(\ref{itse}) and Eq(\ref{deltafunc}), $c_i$ is proportional to the signed number of walkers, $N_i$, on determinant $\ket{D_i}$. The walker population dynamics is governed by:
 \beq
  -\partial_t N_i = (H_{ii}-(E_{HF}+S)) N_i + \sum_{j\ne i} H_{ij} N_j  \label{fciqmceq}
 \eeq
where $S$ is the applied shift, which at convergence (keeping the number of walkers fixed) equals the exact correlation energy.
The population dynamics is implemented via the three FCIQMC steps of spawning, diagonal death and walker annihilation. For more details the reader is referred to \cite{original-fciqmc}. 
In practice, a time-average can be taken in the long-time limit, 
so that $c_i\propto \langle N_i\rangle_{t}$.
  
 In the initiator method, the master equation is modified as follows. 
 \begin{widetext}
 \beq
  -\partial_t N_i = (H_{ii}-(E_{HF}+S)) N_i + \sum_{j\ne i} \tilde{H}_{ij} (N_i,N_j) N_j 
 \eeq        
 \end{widetext}
 where 
 \beq
 \tilde{H}_{ij} (N_i,N_j) = 
 \begin{cases}
 H_{ij} \text{ if } |N_j|> n_a \text{ or } N_i\ne0 \\
 0    \text{ otherwise}
 \end{cases}
 \eeq
 $\tilde{H}$ is the population-dependent truncated Hamiltonian.
 In practice, the initiator algorithm is implemented as follows: for an non-initiator determinant, say $i$, an attempt is made to spawn onto {\em any} of its connected determinants, say $j$, with probability proportional to $H_{ij}$. {\em If $j$ is found to be empty 
the move is rejected}. The initiator rule therefore suppresses spawning events from low-amplitude determinants (the non-initiators) onto empty sites. If these are not suppressed (as in the full method), it is found that there is an extremely rapid, exponential, increase in walker population which is difficult to control (until the annihilation events become sufficiently
frequent to counter this rapid exponential growth). 
It is important to note that all spawns onto occupied sites are however allowed, and therefore the initiator modification to the Hamiltonian is quite subtle and dynamic: as the  
number of walkers increases, an increasing amount of the Hilbert space becomes populated, and as a result there are fewer initiator-rule rejections. On the other hand, for not very large walker populations, it is typically found that the majority of non-initiator spawns are disallowed (rejected Monte Carlo moves), because, for a typical non-initiator, the number of occupied neighbouring determinants (its local Hilbert space) is quite sparsely populated. As a result, many Hamiltonian matrix elements belonging to a non-initiator determinant are effectively zeroed, meaning that 
the local Hilbert space is underpopulated, as compared to what it would be if the fully unconstrained
Hamiltonian were to be applied. 
This leads to an under-sampling bias, since the  {\em feedback} from the local Hilbert space onto the determinant is also smaller than it should be. 
It is this bias that we wish to rectify.  
 
To account for this bias, we now modify the shift applied to a non-initiator determinant, such that instead of applying the full shift, $S$, we apply instead a {\em local} 
shift $S_i$, appropriate for that determinant: 
\beq
S_i=S\times p_{acc}[D_i]
\eeq 
where $p_{acc}[D_i]$ is measured in the simulation itself by monitoring the fraction of spawns
from $\ket{D_i}$ that have been accepted or rejected owing to the initiator rule. 
In other words, the master equation for a non-initiator is modified as:
\begin{widetext}
\beq
  -\partial_t N_i = (H_{ii}-(E_{HF}+S_i)) N_i + \sum_{j\ne i} \tilde{H}_{ij}(N_i,N_j) N_j 
 \eeq        
 \end{widetext}
This equation defines the {\em adaptive-shift} method, in which the shift being applied to non-initiators is modified (reduced) according the rejection probabilities of attempted spawns.  
In order to obtain $p_{acc}[D_i]$, we accumulate two sums, over the accepted ($A_i$) and rejected ($R_i$) spawns respectively, from $D_i$:
\beq
  A_i & = & \sum_{j\in accepted} w_{ij} \\
  R_i & = & \sum_{j\in rejected} w_{ij} \\
  p_{acc}[D_i]&=&\frac{A_i}{A_i+R_i}
\eeq
$w_{ij}$ is a weight to be assigned for each attempted $i\rightarrow j$ spawn, 
whose form will be derived shortly from perturbation theory. 
For the moment let us note
that if the determinant $\ket{D_i}$ becomes an initiator, the full shift is applied, since in that case 
$R_i=0$ and therefore  $p_{acc}[D_i]=1$. Similarly, as the number of walkers $N_w$ increases, the local Hilbert space surrounding $D_i$ becomes populated and in that case also $p_{acc}\rightarrow 1$, i.e. 
 \begin{equation}
 \lim_{N_w \rightarrow\infty} p_{acc}[D_i]=1 \mbox{ for all } i   \label{limpacc}
 \end{equation}  
The full master equation of FCIQMC is therefore re-obtained in the large walker limit.  

The simplest choice to make for the weights, $w_{ij}$ is to set them all to be unity. This choice is actually 
acceptable, in that it ensures the crucial limit of Eq.~(\ref{limpacc}). However it ignores the fact that not all
determinants $j$ in the local Hilbert space of $i$ can be expected to have equal weight, especially in ab initio
Hamiltonians where the different $j$ can be coupled to $i$ with non-uniform matrix element magnitudes $|H_{ij}|$, 
and also can have strongly varying local energies $H_{jj}$.  To take into account the expected non-uniformity in the importance of the determinants in the local Hilbert space of $i$, we can appeal to the concepts behind L\"{o}wdin downfolding \cite{Lowdin1951}. In that procedure, if a determinant $j$ is to be downfolded into $i$, then the off-diagonal Hamiltonian matrix element $H_{ij}$ is zeroed, and 
the diagonal matrix element $H_{ii}$ is changed 
by $H_{ij}^2/(E_0-H_{jj})$, where $E_0$ is the exact energy. This therefore constitutes an 
effective reduction in the energy of $i$. This is a second-order perturbation theory argument: $H_{ij}/(E_0-H_{jj})$ represents the first-order perturbative amplitude of 
$D_j$ due to $D_i$, and the further factor of $H_{ij}$ represents
the feed-back from $D_j$ to $D_i$. The overall effect is then well-known perturbation expression. Garniron et al \cite{Garniron2018} have recently used a similar argument to dress effective Hamiltonians obtained within a CIPSI selected-CI approach.

Motivated by this perturbation theory argument, we define $w_{ij}$, the weight assigned to attempted spawn on $D_j$ from $D_i$, to be:
\beq
  w_{ij} = \frac{|H_{ij}|}{H_{jj}-E_0}
\eeq
Physically this is the neglected amplitude on $D_j$ due to a walker on $D_i$ based on perturbation theory.  Although as an estimate of the neglected amplitude this cannot be exact, in practice the errors made by the perturbation theory estimate are largely inconsequential, since the unbiasing procedure is constructed to become small and eventually disappear 
in the large walker limit. 

From an implementation point of view, this form of $w_{ij}$ imposes a small
overhead compared to a standard initiator algorithm, since the energy $H_{jj}$ of determinant $D_j$ of an attempted spawn must be calculated even if the move
is going to be rejected (in the standard algorithm $H_{jj}$ is calculated only if the spawn onto $D_j$ is accepted). However this overhead turns out be a negligible compared with the benefits of the methodology in terms of speed of convergence 
with respect to walker number. %Everytime a spawn is attempted from $D_i$, the quantities $A_i$ and $R_i$ are updated

\section{Results}

\subsection{Butadiene}
We first apply the adaptive-shift FCIQMC method to the  butadiene molecule in an ANO-L-pVDZ basis (22 electrons in 82 orbitals), which has proven challenging for the normal initiator method. For example, 
initiator calculations with $10^9$ walkers with the conventional method 
yielded an energy of $-155.5491(4)$ a.u.\cite{Daday2012}, which is some 8 mH above a  DMRG calculation ($-155.5573$) obtained with a large (6000) number of renormalised functions \cite{Olivares2015}. 
Whilst the exact energy for this system is not known, it can be expected to be only slightly lower than this, most likely within a mH, and other highly accurate methods are consistent with this:  
CCSDT(Q) yields $-155.55756$ and CCSDTQ $-155.55738$, whilst extrapolated HCIPT2 yields $-155.5582(1)$ \cite{Chien2018}.

For the present study, as in the original study \cite{Daday2012}, restricted Hartree-Fock orbitals were used. The calculations were done by starting 100 walkers on the HF determinant, and growing the population (using $S=0$) until $N_w$ reached the target
population, after which the shift was allowed to vary, in adaptive shift mode. The target populations were set to 10M, 50M, 100M and 200M walkers. (In the final case, the 200M walker simulations were grown from the  equilibrated 100M walker simulations). In order to assess the dependence of the results on the 
initiator parameter $n_a$, three independents sets of calculations were performed, using $n_a=3,10, 20$.  The calculations were run for 50000 time-steps,
using a time set of $\Delta \tau=10^{-3}$ a.u. We used a semi-stochastic space \cite{Umrigar2012} of $|{\cal D}|=10^3 (10^4)$ for the systems up to 100M (200M) walkers, selected from the most populated determinants 1000 time-steps after variable shift mode had been reached, this being the (`pops-core') protocol in the NECI code \cite{Blunt2015}. A trial wavefunction space of ${\cal |T|}=10$ of the leading determinants was also used to compute projected energies. For systems dominated by one determinant as in the present case, the results obtained from projections on the Hartree-Fock determinant and the trial wavefunction are found to be very similar. For example, at 200M walkers with $n_a=10$, the trial wavefunction and the HF projected energy both yield an energy of $-155.5583(2)$, i.e. in agreement down to a stochastic error of 0.2 mH. Similar (albeit very slightly worse) agreement is found in the smaller, 10M walker, simulation ($-155.5523(3)$ and $-155.5524(4)$ for the trial and HF-projected wavefunctions respectively). For consistency, all results reported below will be based on the trial wavefunction projected energies.         

As a control, a similar set of calculations were run in the normal initiator mode (at target populations 10M, 100M and 200M), with the three values of initiator parameter.

\begin{figure*}
% \centering
 \includegraphics[width=\textwidth]{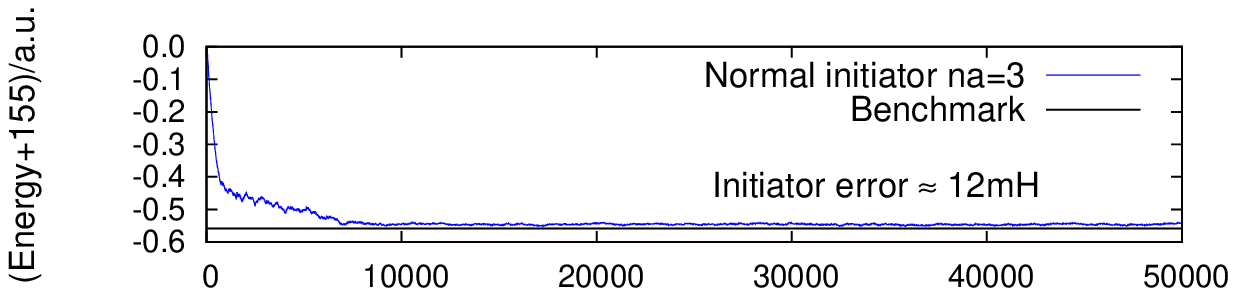}
 
 \includegraphics[width=\textwidth]{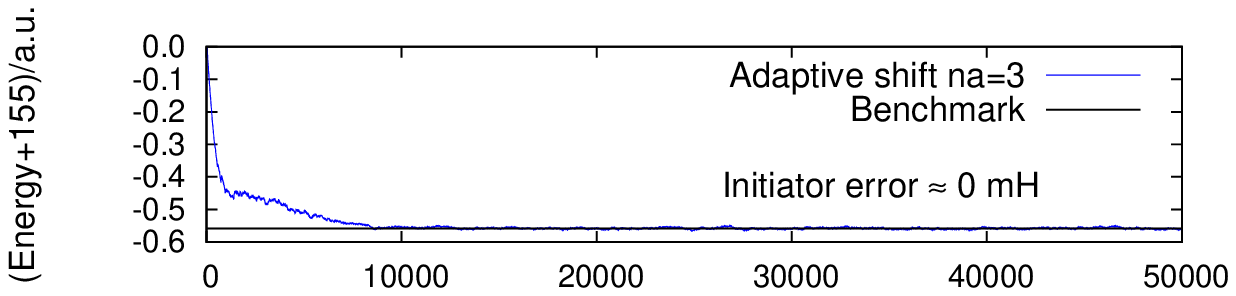}
 
 \includegraphics[width=\textwidth]{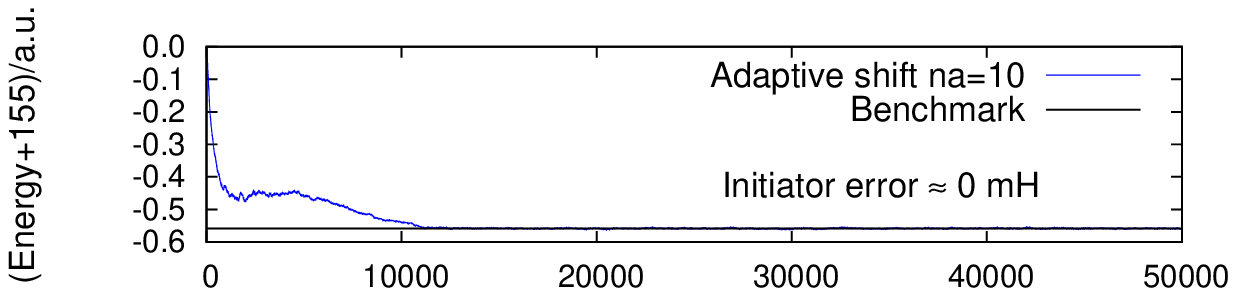}
 
  \includegraphics[width=\textwidth]{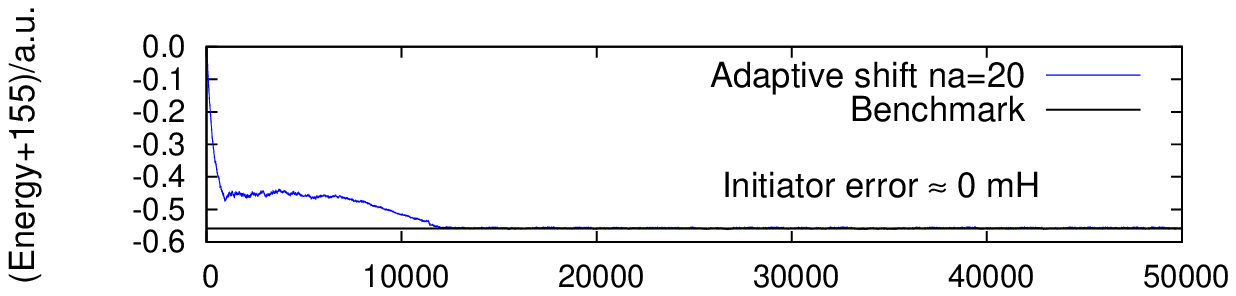}
 
 \includegraphics[width=\textwidth]{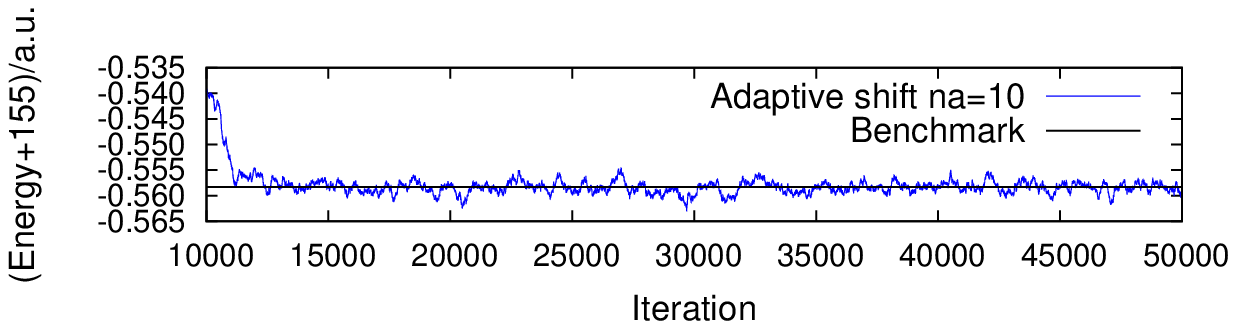}
 \caption{Total energy trajectories of 100M walker FCIQMC simulations of butadiene/ANO-L-pVDZ, using the standard initiator (top panel) and adaptive shift method (2nd, 3rd and 4th panels with $n_a=3, 10, 20$ respectively).The 5th panel is a zoom-in of the $n_a=10$ simulation. The extrapolated HCIPT2 result of $-155.5582$ is used as the benchmark.}
 \label{fig:butaruns}
 \end{figure*}

Trajectories of the calculations are shown in Fig.~\ref{fig:butaruns}, for the standard initiator and adaptive shift methods, for different values of $n_a$. 
In all cases, the simulations converge from the Hartree-Fock determinant to their equilibrium, steady-state, distribution within $\sim 10000$ iterations (i.e. 10 a.u.
of imaginary time), and are thereafter stable, exhibiting small fluctuations of a few mH. However, it is evident that the standard initiator method incurs a noticeable bias relative to the 
benchmark, whereas the three adaptive runs, with the very different values of the initiator threshold, all agree extremely well with the benchmark. The larger values 
of $n_a$ tend to exhibit smaller fluctuations in the projected energy. This is because with the larger values of $n_a$, the reference (HF) population, as well as those of the singles and doubles, tends to be higher than when 
small $n_a$ is used, leading to smaller fluctuations in the projected energy.

\begin{figure*}
\centering
\includegraphics[width=300pt]{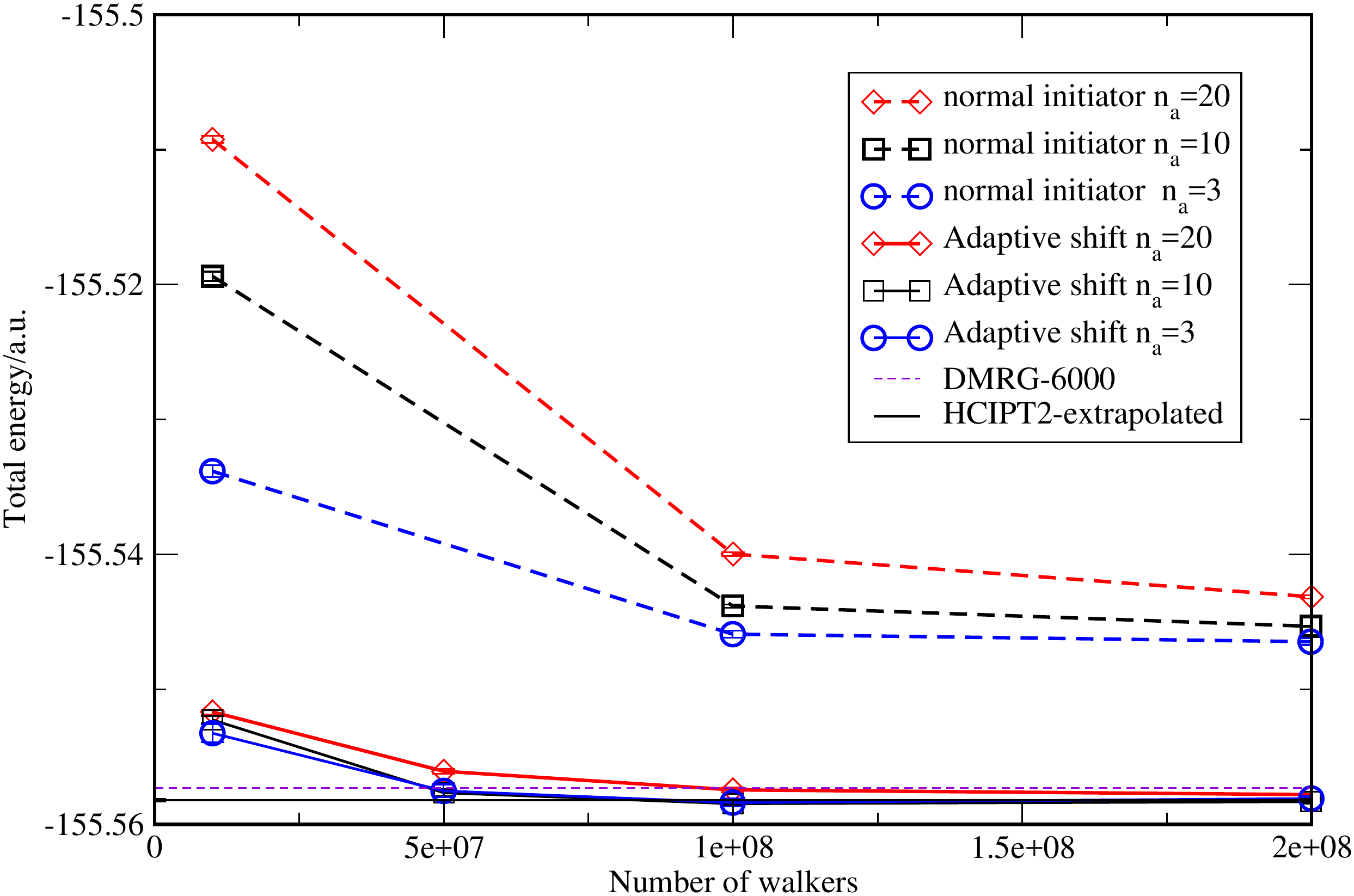}
\caption{Total energies of butadiene/ANO-L-pVDZ for both normal initiator and adaptive shift methods, as a function of the number of walkers, for three values of the initiator parameter $n_a$.}
\label{fig:buta}
\end{figure*}

The full results of the butadiene simulations are shown in Table~\ref{tab:buta}, and in Fig.~\ref{fig:buta}. It is clear that the adaptive shift simulations, irrespective of the value of the initiator parameter
 used, converge to a narrow range of energies, ranging from $-155.5578(2)$ to $-155.5583(2)$, which are in very good agreement with the benchmark value. The fact that the result is
largely independent of the initiator value is remarkable: the different values of the initiator parameter lead to calculations with very different number of initiators in the simulations:
for example, at 200M walkers the $n_a=3$ simulation has $6\times10^5$ initiator determinants, whilst the $n_a=20$ simulation has $5\times10^4$, an order of magnitude fewer. Yet the fact that 
the projected energies are essentially independent of this implies that the adaptive shift method is correctly removing the under-sampling bias of each non-initiator, so that the {\em ratio} of the amplitude of a given non-initiator to the reference determinant is correct, this being the necessary requirement to obtain the exact energy.

\begin{table*}[t]
\begin{center}
{\footnotesize
\begin{tabular}{@{\extracolsep{4pt}}rlll| lll@{}}
\hline
                                &   \multicolumn{3}{c|}{Initiator}                               &                        \multicolumn{3}{c}{Adaptive shift }           \\  
   $N_w/10^6$        &  $n_a=3$                         &    $n_a=10$                &    $n_a=20$    &    $n_a=3$               &      $n_a=10$                         &     $n_a=20$                                         \\ \hline
   $10$                   &   $ -.5338(4)$                    &   $-.5194(3) $             &    $-.5092(3)$  &    $ -.5532(7)$          &     $-.5523(3) $                     &      $-.5517(2)$             \\
  50                        &                                           &                                   &                              &    $-.5575(5) $          &     $-.5577(3) $		        &      $-.5555(2)$      \\ 
  100                      &        $-.5459(2)$                &   $-.5438(1)$              &    $-.5399(2)$  &    $-.5583(3) $         &      $-.5584(2) $			&	$-.5574(2)$	 \\ 
  200                      &       $-.5465(2)$                 &   $-.5453(1) $             &    $-.5431(1)$  &    $-.5581(2) $	        &     $-.5583(2) $			&	 $-.5578(2)$ \\
           \hline
  CCSD(T)          &&&  &   && $-.5550  $        \\ 
  CCSDT(Q)        &&&  &   & &  $-.5576  $     \\
  DMRG-6000      &&&    &  & & $-.5573 $    \\
  HCIPT2 (extrap)   &&&  & & & $-.5582(1) $    \\
\hline

\end{tabular}
}
\caption{Butadiene total energy in a.u. (offset by 155~a.u.) }
\label{tab:buta}
\end{center}
\end{table*}

\subsection{Benzene} 

\begin{figure*}
 \centering
 \includegraphics[width=\textwidth]{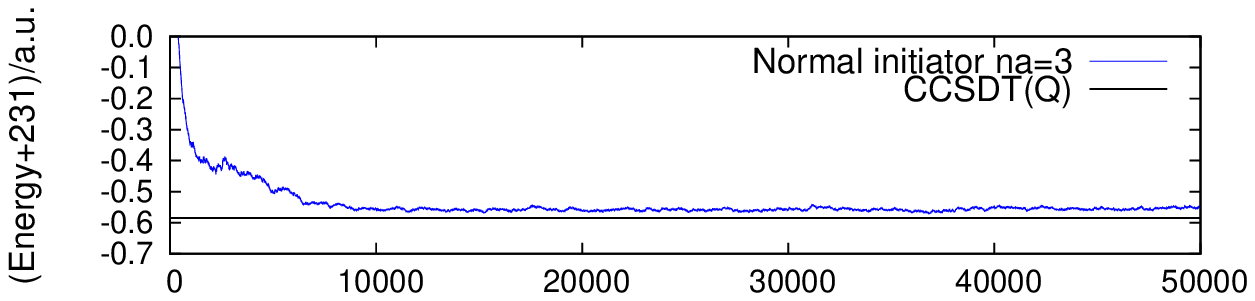}
 
 \includegraphics[width=\textwidth]{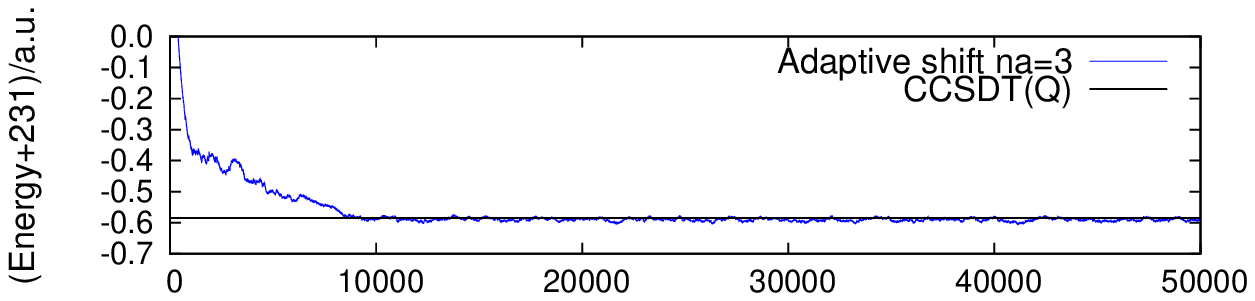}
 
 \includegraphics[width=\textwidth]{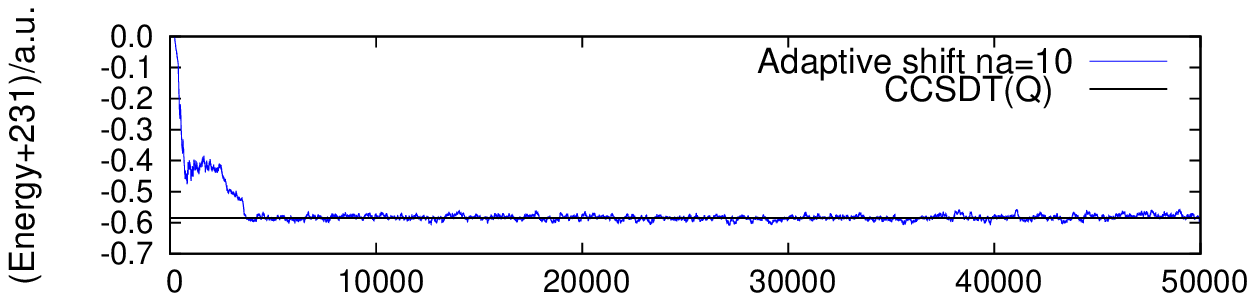}
 
  \includegraphics[width=\textwidth]{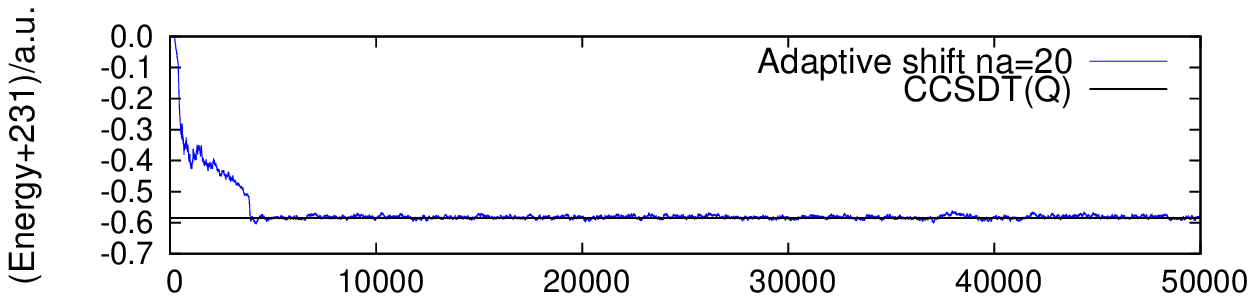}
 
 \includegraphics[width=\textwidth]{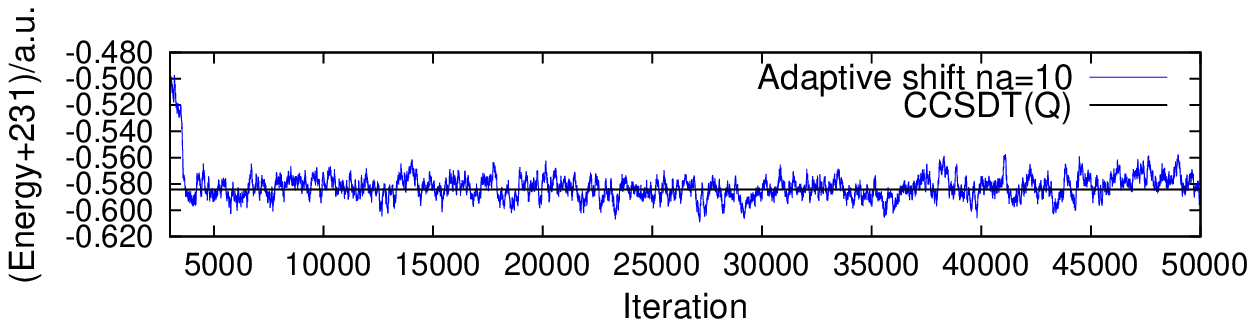}
 \caption{Total energy trajectories of 100M walker FCIQMC simulations of benzene/cc-pVDZ, using the standard initiator (top panel) and adaptive shift method (2nd, 3rd and 4th panels with $n_a=3, 10, 20$ respectively). The 5th panel is a zoom-in of the $n_a=10$ simulation. }
 \label{fig:benzeneruns}
 \end{figure*}

\begin{figure*}
\centering
\includegraphics[width=300pt]{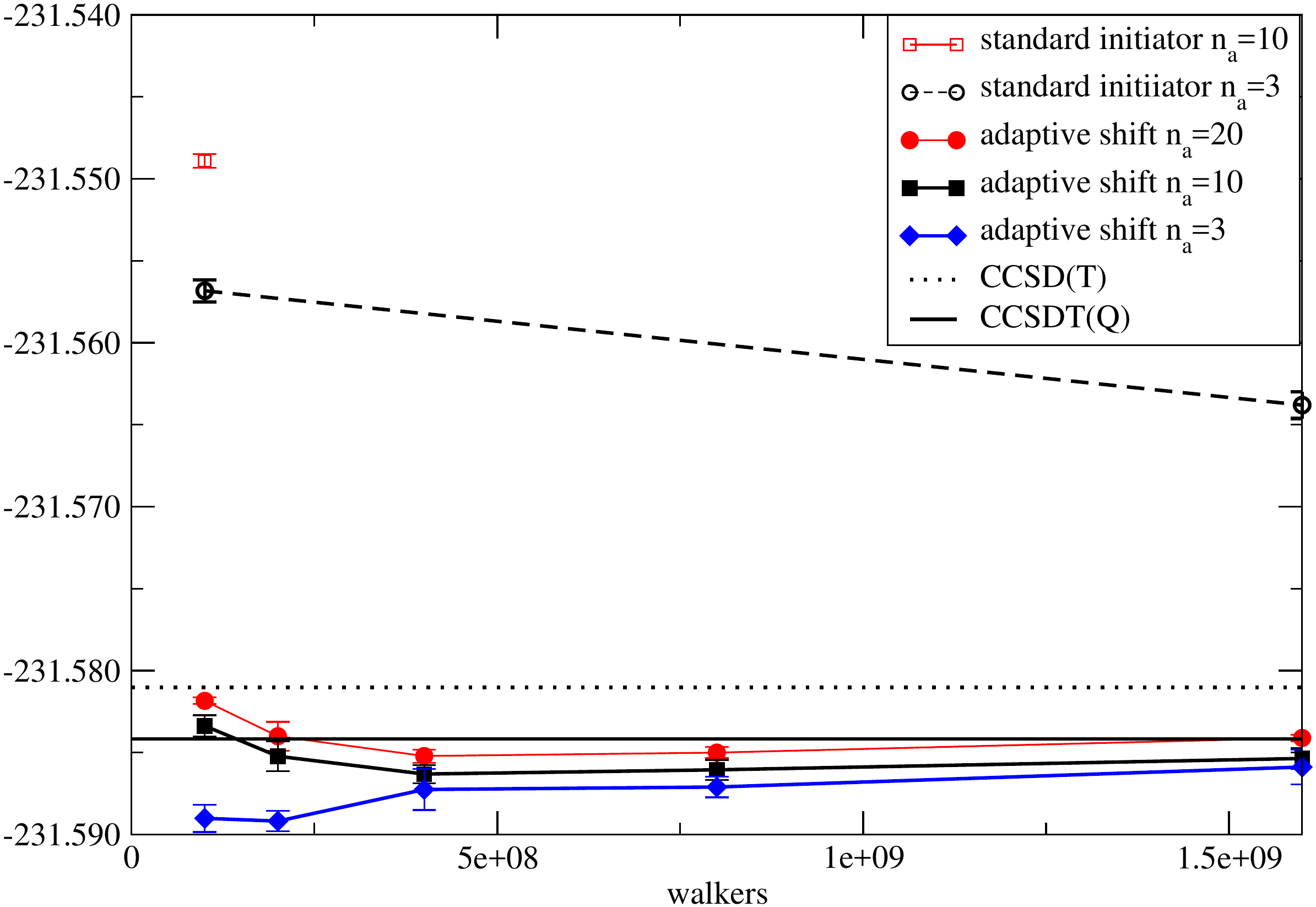}
\caption{Total energy of benzene/cc-pVDZ with walker number for different initiator parameters, for the standard initiator and adaptive shift methods}
\label{fig:benzene}
\end{figure*}

Next we report adaptive shift calculations on the ground state of the benzene molecule in a cc-pVDZ basis (30 electrons in 108 orbitals), at the experimental geometry given on the NIST website (see Supplemental Material).  In a $D_{2h}$ point group, the Hilbert space 
is $\sim 10^{35}$. CASSCF(6,12) orbitals were used, as in a previous study using linearised coupled cluster theory \cite{Jeanmairet2017}. Similar to the butadiene calculations, calculations were performed at three values of the initiator parameter, $n_a=3, 10, 20$, with walkers in the range $100$M to 1.6B. The semi-stochastic and trial-wavefunction spaces were also similarly chosen. Trajectories of the 100M walkers simulations 
are shown in Fig.~\ref{fig:benzeneruns}. The behaviour
observed for this much larger system is similar to that of butadiene, with a noticeable initiator bias in the 
normal initiator method ($\sim 20$ mH at 1.6B walkers at $n_a=3$), and a much reduced error in the adaptive shift simulations. 
The complete results are shown in Fig.~\ref{fig:benzene}. It is seen that even with $10^8$ walkers, excellent energies are obtained, 
$-231.589(1)$ at $n_a=3$, $-231.5833(7) (n_a=10)$, and $-231.5818(2) (n_a=20)$, to be compared with the CCSDT(Q) value ( $-231.58416$). Compared to the butadiene simulation, the fluctuations in the instantaneous projected energy are somewhat larger, about 20 mH rather than 5 mH, but given the much larger size of the problem with increased connectivity around each determinant of a factor of $\sim 4$, this is not 
surprising. As the walker
number is increased to 1.6B, these numbers converge into a narrower range of less than 1.6 mH: $-231.5858(2)$ ($n_a=3$), $-231.5853(6)$ ($n_a=10$), $-231.5841(3)$ ($n_a=20$). The main difference observed here, compared to butadiene, is that the $n_a=3$ simulation converges from below, with an overshoot of 
about 4~mH before rising to the above value. The two larger initiator parameters potentially also exhibit overshoots, but these are much smaller, 
about 1~mH, and well within within the stochastic fluctuations of simulation, as demonstrated in the zoom-in of the $n_a=10$ simulation Fig.(\ref{fig:benzeneruns}). Overall, it is difficult to pinpoint the energy with higher accuracy than 1 mH, and we would suggest that the exact 
answer lies within a mH of $-231.585$, which is consistent with the CCSDT(Q) energy of  $-231.58416$.. Normally
for such systems dominated by dynamical correlation, the CC hierarchy converges from above, and our best estimate of 
$-231.585$ is indeed slightly below the CCSDT(Q) value. 

%\begin{table*}[t]
%\begin{center}
%{\footnotesize
%\begin{tabular}{@{\extracolsep{4pt}}r|lll@{}}
%\hline
%   $N_w/10^6$  &  $n_a=3$                  &  $n_a=10$           &     $n_a=20$                                         \\ \hline
%  $100$        &  -.5890(9)                 &        $-.5833(7)$  &       $-.5818(2)   $                                              \\ 
%  $200$        &  -.5891(7)                 &        $-.5852(9)$   &      $-.5840(8) 	$					\\                 
%  $400$        &  -.5872(2)                 &        $-.5863(6)$  &       $-.5852(4) 	$				\\
%  $800$        &  -.5870(6)                 &        $-.5860(7)$  &       $-.5850(3) 	$				\\
%  $1600$       &  -.5857(2)                 &        $-.5853(6)$  &       $-.5842(3)$   					\\  \hline
%  CCSD(T)   &     &    &    $ -.5810$  \\ 
%  CCSDT(Q)  &         &     &   $ -.5842$ \\
%\hline
%\end{tabular}
%}
%\caption{Benzene total energy (offset by 231) with the adaptive shift method.}
%\label{tab:benzene}
%\end{center}
%\end{table*}

\section{Concluding remarks}
In conclusion, we have demonstrated an adaptive-shift method to unbias the initiator bias on the fly in an i-FCIQMC calculation, resulting in highly accurate simulations of sizeable systems such as benzene/cc-pVDZ.  Near-FCI quality energies can be obtained with drastically reduced number of walkers as compared to the standard initiator method. The internal consistency of the methodology is demonstrated by the fact that the dependence of the method on the initiator parameter is largely removed, enabling converged results to be obtained even with large values of the initiator parameter. The advantage of using a large initiator parameter is that the reference population is much larger, leading to smaller fluctuations in the simulations. The latter will prove very useful in multireference systems, where populations on the reference determinants tend to be small, and require large initiator thresholds for stabilisation. The fact that we can now correctly unbias the simulations even when the initiator threshold is large will be extremely beneficial in the treatment of strongly correlated, multireference systems, which we will return to in subsequent work. In addition, in contrast to methods which rely on extrapolations to the FCI limit to achieve accuracy, the present method yields near-exact density matrices, which can be used to calculate properties. This will be the subject of a forthcoming publication. 

\section{Supplementary Material}
The geometry of the benzene molecule used in this study is specified in the Supplementary material file. 

\section{Acknowledgements}
The authors gratefully acknowledge funding from the Max Planck Society.

\newpage
\bibliography{adaptiveshift.bib}
                     
\end{document}

\maketitle

Geometry of benzene used in this study (\AA ngstr\"{o}m)

\begin{verbatim}
*  Expt geometry from http://cccbdb.nist.gov/exp2.asp 
C 	0.0000	1.3970	0.0000
C 	1.2098	0.6985	0.0000
C 	1.2098	-0.6985	0.0000
C 	0.0000	-1.3970	0.0000
C 	-1.2098	-0.6985	0.0000
C 	-1.2098	0.6985	0.0000
H 	0.0000	2.4810	0.0000
H 	2.1486	1.2405	0.0000
H 	2.1486	-1.2405	0.0000
H  	0.0000	-2.4810	0.0000
H  	-2.1486	-1.2405	0.0000
H  	-2.1486	1.2405	0.0000
\end{verbatim}